\documentclass[lettersize,journal]{IEEEtran}
\makeatletter
\def\bstctlcite{\@ifnextchar[{\@bstctlcite}{\@bstctlcite[@auxout]}}
\def\@bstctlcite[#1]#2{%
 \@bsphack
 \@for\@citeb:=#2\do{%
 \edef\@citeb{\expandafter\@firstofone\@citeb}%
 \if@filesw\immediate\write\csname #1\endcsname{\string\citation{\@citeb}}\fi}%
 \@esphack}
\makeatother
\IEEEoverridecommandlockouts
\usepackage[T1]{fontenc}
\usepackage[english]{babel}
\usepackage{cite}
\ifCLASSINFOpdf
  \usepackage[pdftex]{graphicx}
\fi
\usepackage{amsmath}
\usepackage{amsmath,amssymb,amsfonts}
\usepackage{algorithm}
\usepackage{algorithmic}
\usepackage{array}
\usepackage{float}
\usepackage[caption=false,font=footnotesize]{subfig}
\usepackage{stfloats}
\usepackage{url}
\newcommand{\Matlab}{\textsc{Matlab}}
\usepackage{makecell}

\newcolumntype{P}[1]{>{\centering\arraybackslash}p{#1}}

\newcommand{\bmat}{\begin{bmatrix}}
\newcommand{\emat}{\end{bmatrix}}

\newcommand{\ones}{\mathbf 1}

\newcommand{\vect}[1]{\boldsymbol{#1}}

\newcommand{\norm}[1]{\lVert #1 \rVert}

\newcommand{\vW}{{\vect W}}

\usepackage{tikz}
\usetikzlibrary{fit}
\usepackage{ellipsis}
\usetikzlibrary{calc}
\usetikzlibrary{decorations.pathreplacing,decorations.markings,shapes.geometric}
\usetikzlibrary{chains}
\tikzstyle{block} = [draw, rectangle, minimum height=3em, minimum width=4.5em]
\tikzstyle{input} = [coordinate]
\tikzstyle{output} = [coordinate]
\tikzstyle{pinstyle} = [pin edge={to-,thin,black}]

\usetikzlibrary{positioning}

\tikzset{radiation/.style={{decorate,decoration={expanding waves,angle=90,segment length=5pt}}}}
\tikzset{font={\fontsize{10pt}{12}\selectfont}}

\usetikzlibrary{spy}
\usepackage{pgfplots}
\usepackage{wrapfig}
\usetikzlibrary{arrows,shapes}
\usetikzlibrary{positioning,shapes.callouts}
\usepgfplotslibrary{groupplots,dateplot}
\usetikzlibrary{patterns,shapes.arrows}
\pgfplotsset{compat=newest}
\def\BibTeX{{\rm B\kern-.05em{\sc i\kern-.025em b}\kern-.08em
    T\kern-.1667em\lower.7ex\hbox{E}\kern-.125emX}}

\definecolor{TUMBeamerYellow}    {rgb} {1.000,0.706,0.000}    
\definecolor{TUMBeamerOrange}    {rgb} {1.000,0.502,0.000}    
\definecolor{TUMBeamerRed}       {rgb} {0.898,0.204,0.094}    
\definecolor{TUMBeamerDarkRed}   {rgb} {0.792,0.129,0.247}    
\definecolor{TUMBeamerBlue}      {rgb} {0.000,0.600,1.000}    
\definecolor{TUMBeamerLightBlue} {rgb} {0.255,0.745,1.000}    
\definecolor{TUMBeamerGreen}     {rgb} {0.569,0.675,0.420}    
\definecolor{TUMBeamerLightGreen}{rgb} {0.710,0.792,0.510}    
\definecolor{color2}{rgb}{0.843137254901961,0.0980392156862745,0.109803921568627}
\definecolor{color1}{RGB}{0,101,189}

\newcommand\copyrighttext{%
	\footnotesize \textcopyright This work has been submitted to the IEEE for possible publication. Copyright may be transferred without notice, after which this version may no longer be accessible.}
\newcommand\copyrightnotice{%
	\begin{tikzpicture}[remember picture,overlay]
	\node[anchor=south,yshift=10pt] at (current page.south) {\fbox{\parbox{\dimexpr\textwidth-\fboxsep-\fboxrule\relax}{\copyrighttext}}};
	\end{tikzpicture}%
}

\begin{document}

\title{Learning Representations for CSI Adaptive Quantization and Feedback}

\author{Valentina Rizzello,~\IEEEmembership{Student Member,~IEEE}, Matteo Nerini,~\IEEEmembership{Student Member,~IEEE},\\ Michael Joham,~\IEEEmembership{Member,~IEEE}, Bruno Clerckx,~\IEEEmembership{Fellow,~IEEE}, and Wolfgang Utschick,~\IEEEmembership{Fellow,~IEEE}
\thanks{V.~Rizzello, M.~Joham and W.~Utschick are with the Professur f{\"u}r
Methoden der Signalverarbeitung, Technische Universit{\"a}t M{\"u}nchen, Munich,
80333, Germany. \{{valentina.rizzello, joham, utschick}\}@tum.de}
\thanks{M.~Nerini and B.~Clerckx are with the Department of Electrical and Electronic Engineering, Imperial College London, London, SW7 2AZ, U.K. \{m.nerini20, b.clerckx\}@imperial.ac.uk}
\thanks{This research was supported by an unrestricted gift from Futurewei Technologies, Inc., Huawei R\&D USA.}
}



\maketitle
\copyrightnotice
\begin{abstract}
In this work, we propose an efficient method for channel state information (CSI) adaptive quantization and feedback in frequency division duplexing (FDD) systems. Existing works mainly focus on the implementation of autoencoder (AE) neural networks (NNs) for CSI compression, and consider straightforward quantization methods, e.g., uniform quantization, which are generally not optimal. 
With this strategy, it is hard to achieve a low reconstruction error, especially, when the available number of bits reserved for the latent space quantization is small.
To address this issue, we recommend two different methods: one based on a post training quantization and the second one in which the codebook is found during the training of the AE. Both strategies achieve better reconstruction accuracy compared to standard quantization techniques. 
%
\end{abstract}


\section{Introduction}
\IEEEPARstart{M}{assive} multiple-input multiple-output (MIMO) is one of the most favourable technologies to increase throughput in 5G and beyond 5G networks. Having a base station (BS) equipped with a large antenna array helps to suppress the multiuser interference and to improve the spectral efficiency~\cite{CLERCKX}. However, in FDD mode, this benefit can be only achieved with an accurate downlink (DL) CSI acquisition at the BS. Therefore, in existing FDD MIMO systems, the mobile terminal (MT) first estimates the DL CSI through a pilot training phase and subsequently reports the estimated DL CSI to the BS. Practical approaches to reduce the feedback overhead include sending the CSI only for a subset of antennas~\cite{magazinejeo}, and codebook based approaches in which the MT sends back to the BS a precoding matrix indicator. On the other hand, in the recent years, artificial intelligence (AI)-based methods have been successively developed for CSI compression and reconstruction. The first example in this direction is represented by CsiNet~\cite{csinet}, where the DL CSI is regarded as an image, and an autoencoder based on a convolutional NN is introduced. Building upon the success of CsiNet, the most recent works~\cite{changeable,binary-net} focus on the design of lightweight NNs that reduce the storage space required at the MT and that achieve better encoding efficiency at the same time. Among the latest advances, a user-centric distributed learning has been proposed in~\cite{gossip-train} to improve the feedback accuracy by preventing overfitting.
In this work, instead we consider the setup of~\cite{vari2021a}, where an AE NN is trained at the BS in a centralized manner using uplink (UL) CSI data, and subsequently the encoder part is offloaded to all the MTs in the cell such that they can compress and feedback the DL CSI. Notably, the training procedure, which requires a high computational effort, is completely done at the BS and not at the MTs. This result is based on the idea proposed in~\cite{wout2022} where it is shown that a NN trained on UL data can generalize to DL data, especially, when the frequency gap between the UL and DL center frequencies is small compared to the UL center frequency. 
Recently, the idea of~\cite{wout2022} has been successfully extended also to ``linear'' architectures, such as the principal component analysis (PCA)-based AE in~\cite{icl_paper}. 
Please, note that in the setups of~\cite{vari2021a,wout2022,icl_paper} it is highly recommended that: \textit{i)} the number of parameter of the encoder NN is small, since the encoder has to be offloaded to the users; \textit{ii)} the compressed CSI that has to be fed back to the BS can be encoded with a small amount of bits, reducing in this way the feedback overhead.

The rest of the paper is organized as follows. In the first part of the paper two quantization strategies for the feedback are proposed: one based on the nested dropout (ND) layer~\cite{RippelGA14} and a second one based on the vector quantizer variational AE (VQ-VAE)~\cite{OordVK17}.
With the first approach, we can define an ordering of the latent space units. Therefore, when only a few bits are available for the latent space quantization, we can distribute more bits to the most important units to achieve high reconstruction accuracy. With the second approach, the codebook is found during the training of the AE and the benefits of vector quantization are exploited. 
In the second part, we focus our attention on the reduction of both the number of parameters to be offloaded to the MT, and the complexity of the deep NN in terms of number of floating point operations ($\#$FLOPs), for the considered AE architecture~\cite{vari2021a}.

\section{Nested Dropout}
\label{sec:nd}
The ND layer has been originally proposed in~\cite{RippelGA14}. With the standard dropout, some elements of the input are randomly zeroed according to a Bernoulli distribution with probability $p$ and the optimization is carried on with respect to the remaining units. In contrast, with ND, given an input space of dimension $M$, and an index $b \sim p_B(\cdot)$, all the units from $b+1$ up to $M$ are dropped. In this way, for each training step we assure that if unit $b$ appears, then also units $1, \dots, b-1$ are kept. Since each unit can always rely on the presence of its predecessor, with the ND, we are implicitly defining an importance ranking over the units $1, \dots, M$. Therefore, the units with smaller indices e.g., $1, 2, 3,$ carry more information than the units with larger indices e.g., $M-2, M-1, M$. 
In~\cite{RippelGA14} it has been shown that the ND applied on a model with a linear (or sigmoidal) encoder, and a linear decoder can recover the PCA solution. Regarding $p_B(\cdot)$ the authors in~\cite{RippelGA14} suggest to use a geometric distribution, i.e.,
\begin{equation}
p_B(b) = (1-p)^{(b-1)}p,
\end{equation}
where $p$ represents the probability of success in each trial, and $b$ is the first occurrence with success.
\\
The main idea is to use the ND layer as the last layer of the encoder, as:
\begin{equation}
  f_{\text{enc}}(\cdot) = f_{\text{ND}}(f_{\boldsymbol{\theta}}(\cdot)),
\end{equation}
where $f_{\text{enc}}(\cdot)$, $f_{\text{ND}}$ and $f_{\boldsymbol{\theta}}$ denote the encoder NN, the ND layer, and the rest of the encoder network parameterized by $\boldsymbol{\theta}$, respectively.
Moreover, as the standard dropout layer, also the ND layer is only active during the training phase. However, the parameter $p$ of the geometric distribution must be chosen carefully, since when $p$ is too small, it can happen that the latest units of the latent space remain completely untrained, and this would reduce to the undesired effect of having a latent space with a smaller dimension. Therefore, it is important that $p$ is chosen small enough to guarantee exactly the same performance that one would have without ND layer, with the advantage of having an ordered latent space. In this work, the value of $p$ has been determined by trial and error.
Once we have defined a ranking over the latent space units, we can assign the bits to these units using the greedy Algorithm~\ref{algo: find_bits} discussed in~\cite{vq_book}.
\begin{algorithm}[t]
\caption{Greedy Bit Allocation Algorithm from~\cite{vq_book}}
\begin{algorithmic}
\STATE \textbf{Step 0.} Initialize the bit allocation to zero, so that $b_i(0) =0$ for each $i$ and $n=0$. Set $s_i(0) = W_i(0)$ as the initial values of demand. 
\STATE \textbf{Step 1.} Find the index $j$ with maximum demand.
\STATE \textbf{Step 2.} Set $b_j(n+1) = b_j(n) + 1$, set $b_i(n+1) = b_i(n)$ for each $i \neq j$, and set $s_i(n + 1) = W_i(b_i(n+1))$.
\STATE \textbf{Step 3.} If $n < B-1$ increment $n$ by $1$ and go to step 1. Otherwise stop.
\end{algorithmic}
\label{algo: find_bits}
\end{algorithm}
In the next subsection, the overall procedure that combines the ND layer with the Algorithm~\ref{algo: find_bits} that guarantees an adaptive quantization of the CSI feedback is described.
\subsection{Overall procedure}
Firstly, we train the AE by choosing the appropriate parameter $p$ for the ND layer, in our case we opted for $p=0.0005$. Then, after training, we consider the UL validation set to find the bit allocation with Algorithm~\ref{algo: find_bits}. 
At \textbf{Step 0.}, we initialize $W_i(0) = \sigma_i^2$, where $\sigma_i^2$ denotes the variance with respect to the UL validation set of the $i$-th latent space unit. Then, at \textbf{Step 2.}, the value of $W_i(b)$ for a certain bit allocation $b$ is calculated using the k-means algorithm~\cite{Bishop07} with $2^b$ clusters. In formulas we have:
\begin{equation}
    W_i(b) = \sum_{k=1}^{2^b}\sum_{x_i \in C_k} (x_i - \mu_k)^2
\end{equation}
where $C_k$ denotes cluster $k$, and $\mu_k$ denotes it's centroid. Specifically, $W_i(b)$ denotes the sum of squared errors caused by the k-means quantization. This approach is extremely efficient as it requires running $B$ times a k-means algorithm based on scalar values, where $B$ is the total number of bits available for feedback quantization.

\begin{figure}[!t]
\centering
  \scalebox{0.80}{\input{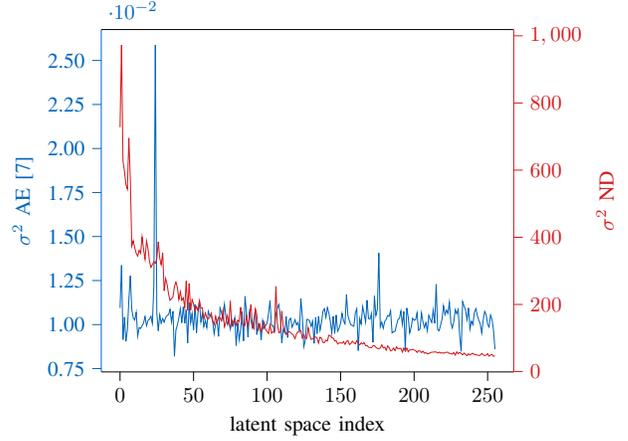}}
  \caption{Variance over the latent space for the UL validation set.}
  \label{fig:variance}
\end{figure}

\section{Vector Quantizer VAE}
\label{sec:vq}
Another approach consists in directly finding the codebook during the training of the NN, rather than after the training has been done.
The vector quantizer variational AE framework (VQ-VAE), which has been originally proposed in~\cite{OordVK17}, can solve this problem. Therefore we decided to use it for our purpose. The codebook for the feedback is learnt during the training of the NN by considering the loss function
\begin{equation}
  \label{eq:loss_vq}
  \mathcal{L}^{\text{vq}} =  \norm{\vect{H} - \hat{\vect{H}}}_{\mathrm{F}}^2 + \norm{\text{sg}[\vect{z}_{\text{e}}] - \vect{z}}_2^2 + \beta \norm{\vect{z}_{\text{e}} - \text{sg}[\vect{z}]}_2^2.
\end{equation}
In particular, we can observe that the loss function in Equation~\eqref{eq:loss_vq} consists of three terms. The first term is the reconstruction loss between the true CSI $\vect{H}$ and the reconstructed CSI at the decoder output $\hat{\vect{H}}$. During the training, the encoder output $\vect{z}_{\text{e}}$ is mapped to the closest codeword $\vect{z}$, which becomes then the input of the decoder. Therefore, the second term, called \textit{quantization loss} optimizes the codebook such that it becomes as close as possible to the encoder output, which is in this case, treated as a constant using the \textit{stop-gradient} operator $\text{sg}[\cdot]$. The last term is the \textit{commitment loss} which in turn optimizes the encoder such that the encoder output $\vect{z}_{\text{e}}$ commits to the codebook. The last term is scaled by a coefficient $\beta < 1$ to make sure that we can still have a good reconstruction. Moreover, because of the quantization in the middle, the training of the encoder parameters through the reconstruction loss is achieved by setting:
\begin{equation}
  \vect{z} \gets \vect{z}_{\text{e}} + \text{sg}[\vect{z} - \vect{z}_\text{e}]
\end{equation}
such that the backpropagation becomes possible. With this approach, the codebook is shared over the latent space units and the codebook size is determined by the length $m$ of the codeword. In particular, it must be that $\mod(M, m)=0$ such that the codebook size is equal to $2^{B/(M/m)}$. Consequently, when $m=1$, we have a scalar quantizer, and when $m>1$, we have a proper vector quantizer with a codebook that consists of different vectors.


\section{Baseline method}
\label{sec:baseline}
As baseline method we consider the AE in~\cite{vari2021a}. However, differently from what has been proposed by the authors, and to further boost the performance, we do not consider a uniform quantization of the latent space but rather a quantization based on the k-means algorithm, where the k-means is run once for each latent space unit. In particular, if $B$ bits are available for the feedback quantization, each element of the latent space is quantized with $B/M$ bits. A similar approach has been considered also in~\cite{icl_paper}. The performance of the baseline method has been  evaluated also with the bit allocation algorithm described in Section~\ref{sec:nd}. However, in this case, the bit allocation algorithm did not improve the result because of the lack of a proper ranking over the latent space units of the AE.

\section{Simulation results - Part 1}
\label{sec:sims1}

\begin{figure}[!t]
\centering
  \scalebox{0.80}{
\begin{tikzpicture}

\definecolor{color0}{rgb}{0.498039215686275,0.788235294117647,0.498039215686275}
\definecolor{color1}{rgb}{0.745098039215686,0.682352941176471,0.831372549019608}
\definecolor{color2}{rgb}{0.992156862745098,0.752941176470588,0.525490196078431}

\begin{axis}[
legend cell align={left},
legend style={fill opacity=1, draw opacity=1, text opacity=1, draw=black},
tick align=outside,
tick pos=left,
x grid style={white!69.0196078431373!black},
xmin=-12.75, xmax=267.75,
xtick style={color=black},
y grid style={white!69.0196078431373!black},
ymin=0.75, ymax=6.25,
ytick style={color=black},
xlabel={latent space index},
ylabel={\# bits},
]
\addplot [semithick, color0, opacity=0.3, mark=*, mark size=2.5, mark options={solid}, only marks]
table {%
0 4
1 4
2 4
3 4
4 3
5 3
6 4
7 3
8 3
9 3
10 3
11 3
12 3
13 3
14 3
15 3
16 3
17 3
18 3
19 3
20 3
21 3
22 3
23 3
24 3
25 3
26 3
27 3
28 3
29 3
30 3
31 3
32 3
33 3
34 3
35 3
36 3
37 3
38 3
39 3
40 3
41 3
42 2
43 3
44 3
45 3
46 2
47 3
48 2
49 3
50 2
51 2
52 2
53 2
54 2
55 2
56 2
57 2
58 2
59 2
60 2
61 2
62 2
63 2
64 2
65 2
66 2
67 2
68 2
69 2
70 2
71 2
72 2
73 2
74 2
75 2
76 2
77 2
78 2
79 2
80 2
81 2
82 2
83 2
84 2
85 2
86 2
87 2
88 2
89 2
90 2
91 2
92 2
93 2
94 2
95 2
96 2
97 2
98 2
99 2
100 2
101 2
102 2
103 2
104 2
105 2
106 3
107 2
108 2
109 2
110 2
111 2
112 2
113 2
114 2
115 2
116 2
117 2
118 2
119 2
120 2
121 2
122 2
123 2
124 2
125 2
126 2
127 2
128 2
129 2
130 2
131 2
132 2
133 2
134 2
135 2
136 2
137 2
138 2
139 2
140 2
141 2
142 2
143 2
144 2
145 2
146 2
147 2
148 2
149 2
150 2
151 2
152 2
153 2
154 2
155 2
156 2
157 2
158 2
159 2
160 2
161 2
162 2
163 2
164 2
165 2
166 2
167 2
168 2
169 2
170 2
171 2
172 2
173 2
174 2
175 2
176 2
177 2
178 2
179 2
180 2
181 2
182 2
183 2
184 2
185 2
186 2
187 1
188 2
189 2
190 2
191 2
192 1
193 2
194 2
195 2
196 1
197 2
198 2
199 2
200 2
201 1
202 1
203 2
204 2
205 1
206 2
207 1
208 1
209 1
210 1
211 1
212 1
213 2
214 1
215 1
216 1
217 2
218 1
219 1
220 1
221 1
222 1
223 1
224 1
225 1
226 1
227 1
228 1
229 1
230 1
231 1
232 1
233 1
234 1
235 1
236 1
237 1
238 1
239 1
240 1
241 1
242 1
243 1
244 1
245 1
246 1
247 1
248 1
249 1
250 1
251 1
252 1
253 1
254 1
255 1
};
\addlegendentry{B=512}
\addplot [semithick, color1, opacity=0.3, mark=square*, mark size=2.5, mark options={solid}, only marks]
table {%
0 4
1 5
2 4
3 4
4 4
5 4
6 4
7 4
8 4
9 4
10 4
11 4
12 4
13 4
14 4
15 4
16 4
17 4
18 4
19 4
20 4
21 4
22 4
23 4
24 4
25 4
26 4
27 4
28 4
29 4
30 4
31 4
32 3
33 4
34 3
35 3
36 3
37 4
38 3
39 3
40 3
41 3
42 3
43 3
44 3
45 4
46 3
47 4
48 3
49 3
50 3
51 3
52 3
53 3
54 3
55 3
56 3
57 3
58 3
59 3
60 3
61 3
62 3
63 3
64 3
65 3
66 3
67 3
68 3
69 3
70 3
71 3
72 3
73 3
74 3
75 3
76 3
77 3
78 3
79 3
80 3
81 3
82 3
83 3
84 3
85 3
86 3
87 3
88 3
89 3
90 3
91 3
92 3
93 3
94 3
95 3
96 3
97 3
98 3
99 3
100 3
101 3
102 3
103 3
104 3
105 3
106 4
107 3
108 3
109 3
110 3
111 3
112 3
113 3
114 3
115 3
116 3
117 3
118 3
119 3
120 3
121 3
122 3
123 3
124 3
125 3
126 3
127 3
128 3
129 3
130 3
131 3
132 3
133 3
134 3
135 3
136 3
137 3
138 3
139 3
140 3
141 3
142 3
143 3
144 3
145 3
146 3
147 3
148 3
149 3
150 3
151 3
152 3
153 3
154 3
155 3
156 3
157 3
158 3
159 3
160 3
161 3
162 3
163 3
164 3
165 3
166 3
167 3
168 3
169 3
170 3
171 3
172 3
173 3
174 3
175 3
176 3
177 3
178 3
179 3
180 3
181 3
182 3
183 3
184 3
185 3
186 3
187 3
188 3
189 3
190 3
191 3
192 2
193 3
194 3
195 3
196 3
197 3
198 3
199 3
200 3
201 2
202 3
203 3
204 3
205 2
206 3
207 2
208 2
209 2
210 2
211 3
212 3
213 3
214 2
215 3
216 3
217 3
218 2
219 2
220 2
221 2
222 3
223 2
224 2
225 3
226 2
227 3
228 2
229 3
230 2
231 3
232 2
233 2
234 2
235 2
236 2
237 2
238 2
239 2
240 3
241 2
242 2
243 2
244 2
245 2
246 3
247 2
248 2
249 2
250 3
251 2
252 2
253 2
254 2
255 2
};
\addlegendentry{B=768}
\addplot [semithick, color2, opacity=0.3, mark=diamond*, mark size=2.5, mark options={solid}, only marks]
table {%
0 5
1 6
2 5
3 5
4 5
5 5
6 5
7 5
8 5
9 5
10 5
11 5
12 5
13 5
14 5
15 5
16 5
17 5
18 5
19 5
20 5
21 5
22 5
23 5
24 5
25 5
26 5
27 5
28 5
29 5
30 5
31 5
32 4
33 5
34 4
35 4
36 4
37 4
38 4
39 4
40 4
41 4
42 4
43 4
44 4
45 5
46 4
47 4
48 4
49 4
50 4
51 4
52 4
53 4
54 4
55 4
56 4
57 4
58 4
59 4
60 4
61 4
62 4
63 4
64 4
65 4
66 4
67 4
68 4
69 4
70 4
71 4
72 4
73 4
74 4
75 4
76 4
77 4
78 4
79 4
80 4
81 4
82 4
83 4
84 4
85 4
86 4
87 4
88 4
89 4
90 4
91 4
92 4
93 4
94 4
95 4
96 4
97 4
98 4
99 4
100 4
101 4
102 4
103 4
104 4
105 4
106 4
107 4
108 4
109 4
110 4
111 4
112 4
113 4
114 4
115 4
116 4
117 4
118 4
119 4
120 4
121 4
122 4
123 4
124 4
125 4
126 4
127 4
128 4
129 4
130 4
131 4
132 4
133 4
134 4
135 4
136 4
137 4
138 4
139 4
140 4
141 4
142 4
143 4
144 4
145 4
146 4
147 4
148 4
149 4
150 4
151 4
152 4
153 4
154 4
155 4
156 4
157 4
158 4
159 4
160 4
161 4
162 4
163 4
164 4
165 4
166 4
167 4
168 4
169 4
170 4
171 4
172 4
173 4
174 4
175 4
176 4
177 4
178 4
179 4
180 4
181 4
182 4
183 4
184 4
185 4
186 4
187 4
188 4
189 4
190 4
191 4
192 4
193 4
194 4
195 4
196 4
197 4
198 4
199 4
200 4
201 4
202 4
203 4
204 4
205 3
206 4
207 3
208 3
209 3
210 4
211 3
212 4
213 4
214 3
215 4
216 4
217 4
218 3
219 3
220 3
221 3
222 4
223 4
224 3
225 4
226 3
227 4
228 3
229 4
230 3
231 4
232 3
233 3
234 3
235 3
236 3
237 3
238 3
239 3
240 4
241 3
242 3
243 3
244 3
245 3
246 4
247 3
248 3
249 3
250 4
251 3
252 3
253 3
254 3
255 3
};
\addlegendentry{B=1024}
\end{axis}

\end{tikzpicture}}
  \caption{Bit allocations for different $B$.}
  \label{fig:bits}
\end{figure}
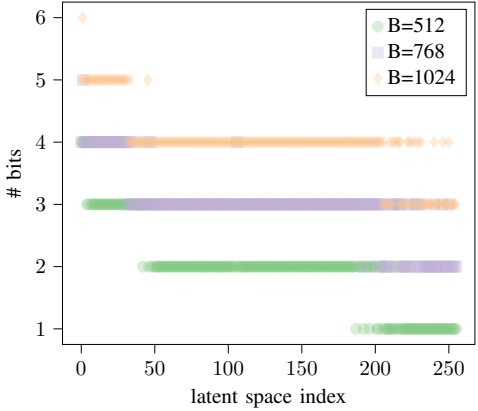

\subsection{Dataset description}
In this work, we consider the same dataset as in~\cite{vari2021a}, where a single urban microcell (UMi) with $150$~meters radius is considered. The CSI datasets at $2.5$~GHz (for UL) and at $2.62$~GHz (for DL) are generated with the $\Matlab$ based software QuaDRiGa version 2.2~\cite{quad}. Specifically, non-line-of-sight (NLoS) channels with $L=58$ micropaths are considered.
The bandwidth is approximately $8$~MHz and is divided over $N_{\text{c}} =160$ subcarriers. The BS is equipped with a uniform planar array with $N_\text{a} = 8\times 8 =64$ antennas, whereas the MTs have a single antenna each.
A total of $60\times 10^3$ samples of both UL CSI, $\vect{H}_{\text{UL}} \in \mathbb{C}^{N_{\text{a}}\times N_{\text{c}}}$, and DL CSI, $\vect{H}_{\text{DL}}$ are collected. Subsequently, the dataset is split into three subsets of training, validation, and testing, each with $48\times 10^3$, $6\times 10^3$, and $6\times 10^3$ samples, respectively.
As outlined before, for training and codebook design only the samples of  $\vect{H}_{\text{UL}} \subseteq \text{\{training, validation\}}$ are used, whereas the performance metrics are evaluated with respect to the samples of $\vect{H}_{\text{DL}} \subseteq \text{\{testing\}}$.
\subsection{Performance metrics}
The performance is evaluated in terms of normalized mean squared error (NMSE)
\begin{equation}
  \varepsilon^2 = \mathbb{E}\left[{\norm{\hat{\vect{H}} - \vect{H}}_{\mathrm{F}}^2}/{\norm{\vect{H}}_{\mathrm{F}}^2}\right]
\end{equation}
where $\vect{H} \in \mathbb{C}^{N_{\text{a}}\times N_{\text{c}}}$ denotes the true channel, and $\hat{\vect{H}} \in \mathbb{C}^{N_{\text{a}}\times N_{\text{c}}}$ denotes the estimated channel at the decoder output.

Additionally, we also evaluate the cosine similarity $\rho$
\begin{equation}
  \rho = \mathbb{E}\left[\frac{1}{N_{\text{c}}}\sum_{n=1}^{N_{\text{c}}}\frac{\vert\hat{\vect{h}}^{\text{H}}_n\vect{h}_n\vert}{\norm{\hat{\vect{h}}_n}_2 \norm{\vect{h}_n}_2}\right]
  \end{equation}
where $\vect{h}_n \in \mathbb{C}^{N_{\text{a}}}$ denotes the true CSI at subcarrier $n$, and $\hat{\vect{h}}_n \in \mathbb{C}^{N_{\text{a}}}$ denotes its estimated version at the decoder output. The cosine similarity measures the quality of beamformer when $\hat{\vect{h}}_n / \norm{\hat{\vect{h}}_n}_2$ is the beamforming vector at the BS.
\subsection{Learning setup}
For all the architectures, we utilize 100 random searches~\cite{random-search} over the most sensitive hyperparameters and we choose the best configuration after 100 epochs. In particular, we tune:
\begin{itemize}
    \item the batch size from $[25, 50, 100]$
    \item the learning rate for the Adam optimizer~\cite{adam} from $ \text{log-unif}(10^{-4}, 10^{-2})$
    \item $\beta$ from $ \text{unif}(0.1, 0.3)$
    \item $m$ from $[1, 2, 4]$.
\end{itemize}
After having chosen the best configuration with the library~\cite{liaw2018tune}, we carry on with the learning until the early stopping criterion with a patience of 30 epochs is met. The early stopping criterion is based on the improvement in the normalized mean squared error of the UL validation set. 
In this section, three different AE framework are considered:
\textit{i)} \textit{``AE''}: the AE in~\cite{vari2021a} presented in Section~\ref{sec:baseline};
\textit{ii)} \textit{``ND''}: a modified version of \textit{``AE''} where the hyperbolic tangent activation function at the encoder output has been replaced by the ND layer as described in Section~\ref{sec:nd};
\textit{iii)} \textit{``VQ-VAE''}: a modified version of \textit{``AE''} where the hyperbolic tangent activation function at the encoder output has been removed and the learning is conduced as described in Section~\ref{sec:vq}.

\subsection{Results}
\label{sec:results}
\begin{figure}[!t]
    \centering
    \includegraphics[scale=0.5]{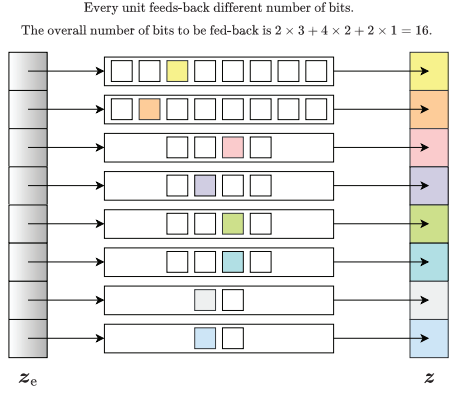}
    \caption{ND based method. Toy example.}
    \label{fig:toy_1}
\end{figure}

\begin{figure}[!t]
    \centering
    \includegraphics[scale=0.4]{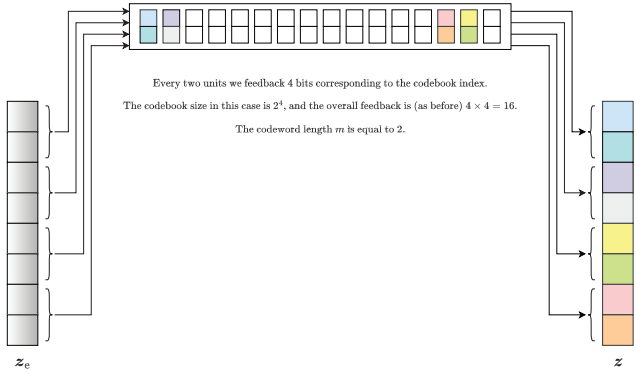}
    \caption{VQ-VAE based method. Toy example.}
    \label{fig:toy_2}
\end{figure}

In Fig.~\ref{fig:variance}, the benefits of the ND layer are highlighted. In particular, we see that all the units in the latent space of \textit{``AE''} carry approximately the same information, whereas for the \textit{``ND''} case, we clearly see that the units corresponding to larger indices have smaller variance compared to the units corresponding to smaller indices. For the sake of completeness, in Fig.~\ref{fig:bits}, we see the results of the bit allocation obtained with Algorithm~\ref{algo: find_bits} for the \textit{``ND''} case. We can observe that the units corresponding to smaller latent space indices receive more bits than the units corresponding to larger indices. In Fig.~\ref{fig:toy_1} and Fig.~\ref{fig:toy_2} two toy examples illustrates how the quantization of the latent vector is performed with the \textit{``ND''} and the \textit{``VQ-VAE''} approach, respectively. For simplicity, in both examples, it is assumed that $M=8$ and $B=16$. In addition, for the \textit{``VQ-VAE''} approach it is further assumed that $m=2$.

\begin{figure}[!t]
\centering
  \scalebox{0.96}{\input{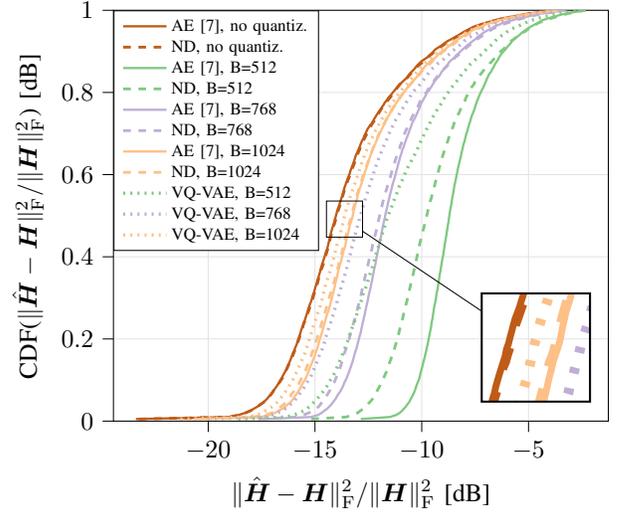}}
  \caption{CDFs of the normalized MSE for DL CSI ($\Delta f=120$~MHz).}
  \label{fig:nmse_g1}
\end{figure}

In Fig.~\ref{fig:nmse_g1}, we show the cumulative distribution functions (CDFs) of the NMSE in decibels for the three different methods taking into account different number of $B$ bits available for the feedback quantization. We can observe that both methods \textit{``ND''} and \textit{``VQ-VAE''} exhibit a gain compared to the \textit{``AE''} method, especially, when the number of available bits $B$ is small, e.g. see $B=512$ in Fig.~\ref{fig:nmse_g1}. Furthermore, we can observe that the \textit{``VQ-VAE''} approach is the one that performs better than all the others. This is due to two reasons: \textit{i)} the quantization is performed during the training, \textit{ii)} vector quantization is used instead of scalar quantization. However, the complexity of the \textit{``VQ-VAE''} method is much larger than the one of the \textit{``ND''} method because of the fact that \textit{i)} the training of a new AE is required every time the number of available bits $B$ changes, and \textit{ii)} because the complexity of vector quantization is higher than the complexity of scalar quantization. Analogous conclusions can be made for the cosine similarity $\rho$. 

\section{Reducing the complexity and the overhead}
\label{sec:compoh}
In this section, we propose some strategies for reducing both the complexity with respect to the of number of FLOPs required by the AE, and the overhead due to the offloading of the encoder parameters from the BS to the MT. Since the \textit{``ND''} method requires the training and consequently the offloading of a single autoencoder architecture, compared to the \textit{``VQ-VAE''} method which requires a different AE depending on the available number of bits $B$, the following discussion focuses on the \textit{``ND''} approach. On the contrary, for both methods, the number of parameters to be offloaded for the codebook is negligible compared to the number of parameters of the encoder.

The complexity of the AE is dominated by the complexity of the convolutional layers, leading to a total of $7.29$ millions FLOPs ($\approx 3.65$ millions MACCs when considering the multiply and add operations as a single instruction) for the encoder network. The reader can refer to~\cite{vari2021a} for the details of the AE architecture, and to~\cite{MolchanovTKAK17} for the details of these computations. In this work, we propose to reduce the complexity of the convolutional layers by replacing convolutional layers with grouped convolutional layers. Hence, we take advantage of the network aggregation principle, as recommended also in~\cite{binary-net}. By assuming that the input of layer $l$ consists of $C^{(l-1)}$ channels and that the output of layer $l$ consists of $C^{(l)}$ channels, we have 
\begin{align}
  \vect{X}_j^{(l)} &= \sum\nolimits_{i=1}^{C^{(l-1)}}\vect{X}_i^{(l-1)} \circledast \vect{K}_{ij}^{(l)} + t_j^{(l)}\ones,\\
  \vect{X}_j^{(l)} &= \sum\nolimits_{i=1}^{G_{n}^{(l-1)}}\vect{X}_i^{(l-1)} \circledast \vect{K}_{ij}^{(l)} + t_j^{(l)}\ones,
\end{align}
for convolutions, and grouped convolutions with $g$ groups, where the $n$-th group considers $G_{n}^{(l-1)} = C^{(l-1)}/g$ channels, respectively. Additionally, it must hold that $\mod{(C^{(l)}, g)}=0$. In particular, $\vect{X}_i^{(l-1)}$ and $\vect{X}_j^{(l)}$, denote the $i$-th input channel and the $j$-th output channel of the $l$-th layer, respectively. $\vect{K}_{ij}^{(l)}$ is the filter that maps $\vect{X}_i^{(l-1)}$ to $\vect{X}_j^{(l)}$ and $t_j^{(l)}$ is the bias term associated with the $j$-th output of layer $l$.
Therefore, it can be deduced that with grouped convolution, the complexity of the convolutional layer is reduced by a factor equal to $g$. 
Regarding the number of parameters to be offloaded to the MT, we observe from~\cite{vari2021a} that the fully connected (FC) layer constitutes the largest overhead with $1280 \times 256 + 256 = 327,936$ single precision floating point numbers, which correspond to $327,936 \times 32$ bits. In order to reduce the amount of bits that have to be offloaded to the UE, we propose two approaches: \textit{i)} replace the weights of the FC layer in the encoder with binary FC weights~\cite{HubaraCSEB16}, \textit{ii)} prune the least significant parameters of the encoder network.


\section{Simulation results - Part 2}
\label{sec:results-two}
\setlength{\arrayrulewidth}{0.1mm}
\renewcommand{\arraystretch}{1.2}

\begin{figure}[!t]
    \centering
    \scalebox{0.8}{
\begin{tikzpicture}

\begin{axis}[
colorbar,
colorbar style={ylabel={}},
colormap={mymap}{[1pt]
  rgb(0pt)=(1,1,0.850980392156863);
  rgb(1pt)=(0.929411764705882,0.972549019607843,0.694117647058824);
  rgb(2pt)=(0.780392156862745,0.913725490196078,0.705882352941177);
  rgb(3pt)=(0.498039215686275,0.803921568627451,0.733333333333333);
  rgb(4pt)=(0.254901960784314,0.713725490196078,0.768627450980392);
  rgb(5pt)=(0.113725490196078,0.568627450980392,0.752941176470588);
  rgb(6pt)=(0.133333333333333,0.368627450980392,0.658823529411765);
  rgb(7pt)=(0.145098039215686,0.203921568627451,0.580392156862745);
  rgb(8pt)=(0.0313725490196078,0.113725490196078,0.345098039215686)
},
xmin=0, xmax=64,
ymin=0, ymax=64,
xtick=\empty, 
ytick=\empty, 
yticklabels={,,},
xticklabels={,,},
]
\addplot [draw=none] coordinates {(0,0)};
\addplot graphics [includegraphics cmd=\pgfimage, xmin=10, xmax=54, ymin=5, ymax=59] {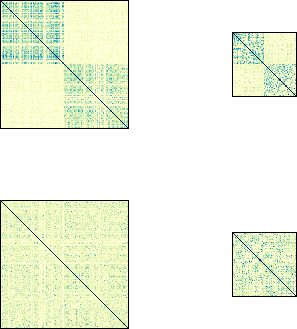};
\end{axis}

\end{tikzpicture}}
    \caption{Heatmaps of the correlation matrix between convolutional filters of two corresponding convolutional layers in \textit{``ND, g=2''} and  \textit{``ND''}, in the 1st and 2nd row, respectively.}
    \label{fig:heatmap}
\end{figure}

In this section, the simulation results related to the approaches described in Section~\ref{sec:compoh} are discussed.
The fourth rows in Tables~\ref{tab1} and~\ref{tab2} show the results in decibels of both the NMSE and the cosine similarity of the DL CSI when using grouped convolutions with two groups $(g=2)$. In this way, the complexity of the encoder decreases to $3.97$ millions FLOPs (1.99 millions MACCs). In particular, when we compare these results with the one obtained for the \textit{``ND''} method, i.e., see the first rows in the Tables, we observe that the grouped convolutions not only reduced the complexity, but also improved the performance metrics. This effect could be explained by the fact that, when $g=2$, we have two parallel architectures, one which focuses on the real part and the other one which focuses on the imaginary part. Therefore, we see an improvement because each filters group is learning a unique representation of the data. As displayed in Fig.~\ref{fig:heatmap}, with grouped convolutions the filters with high correlation are learnt in a more structured way compared to standard convolutions. Additionally, with grouped convolutions, there are less parameters to be learnt in the network, and this can be equivalently interpreted as a regularization effect which prevents the network to overfit.

On the other hand, the remaining rows in Tables~\ref{tab1} and~\ref{tab2} show the results related to the reduction of the overhead due to the offloading of the encoder parameters. In each table, the same colour is utilized to highlight rows with similar results.
\begin{table}[!t]
  \begin{center}
  \caption{NMSE $\varepsilon$~[decibels] with different networks}
  \label{tab1}
    \begin{tabular}{|P{2.44cm}|P{1.2cm}|P{1cm}|P{1cm}|P{1cm}| }
  \hline
  \textbf{NN~($\%$ offload red.)} & \textbf{No~quant.} & \textbf{B=512 }& \textbf{B=768} & \textbf{B=1024}\\
  \hline
  \textcolor{color2}{ND} & \textcolor{color2}{-12.19} & \textcolor{color2}{-8.71} & \textcolor{color2}{-10.64} & \textcolor{color2}{-11.69}\\
  ND, binary FC w. ($\approx 75\%$)& -11.66 & -8.11 & -9.93 & -10.74\\
  \textcolor{color2}{ND, pruning + FT ($\approx 40\%$)} & \textcolor{color2}{-12.12} & \textcolor{color2}{-8.69} & \textcolor{color2}{-10.62} & \textcolor{color2}{-11.62}\\
  \textcolor{color1}{ND, g=2} &\textcolor{color1}{-12.48} & \textcolor{color1}{-8.71} & \textcolor{color1}{-10.78} & \textcolor{color1}{-11.9}\\
  ND, binary FC w., g=2 ($\approx 75\%$) & -11.43 & -8.32 & -10.16 & -11.02\\
  \textcolor{color1}{ND, pruning + FT, g=2 ($\approx 40\%$)} & \textcolor{color1}{-12.45} & \textcolor{color1}{-8.76} & \textcolor{color1}{-10.82} & \textcolor{color1}{-11.88}\\
  \hline 
  ACRNet~\cite{binary-net} &  -10.05 & -- & -- & --\\
  CsiNet~\cite{csinet} & -7.71 & -- & -- & --\\
  \hline
  \end{tabular}
  \end{center}
\end{table}

\begin{table}[!t]
  \begin{center}
  \caption{Cosine similarity $(1 -\rho)$~[decibels] with different networks}
  \label{tab2}
  \begin{tabular}{|P{2.44cm}|P{1.2cm}|P{1cm}|P{1cm}|P{1cm}| }
  \hline
  \textbf{NN~($\%$ offload red.)} & \textbf{No~quant.} & \textbf{B=512 }& \textbf{B=768} & \textbf{B=1024}\\
  \hline
  \textcolor{color2}{ND} & \textcolor{color2}{-14.91} & \textcolor{color2}{-11.55} & \textcolor{color2}{-13.44} & \textcolor{color2}{-14.44}\\
  ND, binary FC w. ($\approx 75\%$)& -14.41 & -11.23 & -13.14 & -14.0\\
  \textcolor{color2}{ND, pruning + FT ($\approx 40\%$)} & \textcolor{color2}{-14.85} & \textcolor{color2}{-11.52} & \textcolor{color2}{-13.41} & \textcolor{color2}{-14.37}\\
  \textcolor{color1}{ND, g=2} &\textcolor{color1}{-15.22} & \textcolor{color1}{-11.56} & \textcolor{color1}{-13.59} & \textcolor{color1}{-14.67}\\
  ND, binary FC w., g=2 ($\approx 75\%$) & -14.19 & -11.16  & -12.97 & -13.8\\
  \textcolor{color1}{ND, pruning + FT, g=2 ($\approx 40\%$)} & \textcolor{color1}{-15.19} & \textcolor{color1}{-11.6} & \textcolor{color1}{-13.62} & \textcolor{color1}{-14.65}\\
  \hline 
  ACRNet~\cite{binary-net} & -13.73 & -- & -- & --\\
  CsiNet~\cite{csinet} & -10.92 & -- & -- & --\\
  \hline
  \end{tabular}
  \end{center}
\end{table}
In particular, for the approach based on binary weights in the FC layer \textit{``binary, FC w.''} we have considered the binarization of the sole weight matrix $\vW \in \mathbb{R}^{d_{\text{out}} \times d_{\text{in}}}$ (the bias vector of the FC layer is therefore represented as single precision floating point number). Moreover, as binarization function we have used the $\mathrm{sign}(\cdot)$ function. Precisely, for the backward pass of the $\mathrm{sign}(\cdot)$ function, the straight-through estimator of the gradient is used, which is defined as $g_{r} = g_{q} 1_{\lvert r\lvert \leq1}$, being $r$ and $q$, the input and output of the $\mathrm{sign}(\cdot)$ function, respectively (see~\cite{HubaraCSEB16}). Furthermore, as recommended in~\cite{RastegariORF16}, we scale the binary weights by  $\alpha=\norm{\vW}_1/(d_{\text{out}} \cdot d_{\text{in}})$. Thus, we can write
$\vect{x}_{\text{out}} = \alpha \vect{W}_b \vect{x}_{\text{in}} + \vect{b}$,
where $\vect{x}_{\text{in}}, \vect{W}_b, \vect{b}, \vect{x}_{\text{out}}$ denote the input of the binary FC layer, the binary weight matrix, the bias vector, and the output of the binary FC layer, respectively. With this approach, we could reduce the number of parameters in terms of bits in the encoder layer by about $75 \%$. Another approach is based on the network pruning and fine tuning, \textit{``pruning + FT''}. More in details, after having chosen the best configuration as described in Section~\ref{sec:results}, we set to zero all the weights with the smallest absolute value, according to a predetermined percentage, e.g., $40\%$. As expected, this degrades the performance. Therefore, after pruning the network, we fine tune the network by optimizing the remaining parameters until the stopping criterion is met. In particular, as highlighted in Tables~\ref{tab1} and~\ref{tab2}, the method \textit{``pruning + FT''} helps to maintain approximately the same performance as the method without pruning and at the same time reduces the number of parameters to be offloaded to the MT by $40\%$. On the other hand, the approach \textit{``binary FC w.''} produces a higher degradation of the performances. Finally, we have compared our method with other two approaches based on AE NNs. In particular, we see that our method outperforms both, the CsiNet~\cite{csinet} and the recent ACRNet presented in~\cite{binary-net}. Differently from our approach, CsiNet and ACRNet have been trained on DL CSI data, and the data have been normalized as specified in the respective works. The two networks leverage the sparsity of the CSI in the delay domain. However, ACRNet boosts the performance by making use of network aggregation in the decoder and by replacing the rectified linear unit (ReLU) activation with a parametric ReLU (PReLU) activation.

\section{Conclusions}
In this work, two different quantization methods based on AE NNs for CSI feedback in FDD systems have been presented. The analysis of the results 
suggests that the reconstruction accuracy can be improved compared to standard quantization approaches. In particular, with the approach based on the ND layer is possible to quantize the latent vector of the AE in an adaptive way by leaving unchanged the AE parameters. Additionally, it has been shown that both the complexity of the AE and the number of parameters of the encoder network can be considerably reduced with the help of grouped convolutions and network pruning, without degrading the overall performance.

\end{document}